 \documentclass[preprint2]{aastex}

\usepackage{lineno}
\usepackage{amsmath,amssymb}
\usepackage{graphicx}
\usepackage{natbib}
\usepackage{hyperref}
\usepackage{color}

\newcommand*\patchAmsMathEnvironmentForLineno[1]{%
  \expandafter\let\csname old#1\expandafter\endcsname\csname #1\endcsname
  \expandafter\let\csname oldend#1\expandafter\endcsname\csname end#1\endcsname
  \renewenvironment{#1}%
     {\linenomath\csname old#1\endcsname}%
     {\csname oldend#1\endcsname\endlinenomath}}%
\newcommand*\patchBothAmsMathEnvironmentsForLineno[1]{%
  \patchAmsMathEnvironmentForLineno{#1}%
  \patchAmsMathEnvironmentForLineno{#1*}}%
\AtBeginDocument{%
\patchBothAmsMathEnvironmentsForLineno{equation}%
\patchBothAmsMathEnvironmentsForLineno{align}%
\patchBothAmsMathEnvironmentsForLineno{flalign}%
\patchBothAmsMathEnvironmentsForLineno{alignat}%
\patchBothAmsMathEnvironmentsForLineno{gather}%
\patchBothAmsMathEnvironmentsForLineno{multline}%
}

\slugcomment{Version 5.0}

\shorttitle{Jupiter as a Giant Cosmic Ray Detector}
\shortauthors{Rimmer, Stark \& Helling}

\begin{document}

\title{Jupiter as a Giant Cosmic Ray Detector}

\author{P. B. Rimmer, C. R. Stark \& Ch. Helling}
\affil{SUPA, School of Physics \& Astronomy, University of St Andrews, North Haugh, St Andrews, KY16 9SS, UK}
\email{pr33@st-andrews.ac.uk}

\begin{abstract}
We explore the feasibility of using the atmosphere of Jupiter to detect Ultra-High-Energy Cosmic Rays (UHECR's). The 
large surface area of Jupiter allows us to probe cosmic rays of higher energies than previously accessible. Cosmic 
ray extensive air showers in Jupiter's atmosphere could in principle be detected by the Large Area Telescope (LAT) 
on the Fermi observatory. In order to be observed, these air showers would need to be oriented toward the Earth, and 
would need to occur sufficiently high in the atmosphere that the gamma rays can penetrate. We demonstrate that, under 
these assumptions, Jupiter provides an effective cosmic ray ``detector'' area of $3.3 \times 10^7$ km$^2$. We 
predict that Fermi-LAT should be able to detect events of energy $>10^{21}$ eV with fluence $10^{-7}$ erg cm$^{-2}$ at a rate 
of about one per month. The observed number of air showers may provide an indirect measure of the flux of cosmic 
rays $\gtrsim 10^{20}$ eV. Extensive air showers also produce a synchrotron signature that may be measurable by ALMA. 
Simultaneous observations of Jupiter with ALMA and Fermi-LAT could be used to provide broad constraints on the energies 
of the initiating cosmic rays.
\end{abstract}

\keywords{cosmic rays --- planets and satellites: individual (Jupiter) --- gamma rays: general --- submillimeter: general}

\section{Introduction}
\label{sec:intro}

When a cosmic ray proton or nucleus of sufficient energy collides with the nucleus of a molecule in Earth's atmosphere, 
the result is a shower of particles spread horizontally over kilometers. For example, a $10^{19}$ eV cosmic ray will 
produce an extensive air shower spread over a few km$^2$ at sea level. These extensive air showers were independently 
observed and identified by \citet{Rossi1934} and \citet{Auger1939}. The relationship between the depth-dependent 
composition and the geometrical extent of extensive air showers as well as the composition and energy of the 
cosmic ray responsible for the primary event can in some cases be determined by reconstructing specific events from 
observation, such as by the Auger Observatory \citep{Abraham2010}. 

The energy of an extensive air shower is defined here as the energy of the particle responsible for the primary event. 
The most energetic extensive air showers detected have energies of $E_0 \sim 10^{20}$ eV. The existence of cosmic 
rays of $\gtrsim 10^{20}$ eV would violate the Greisen Zatsepin Kuzmin (GZK) limit \citep{Greisen1966,Zatsepin1966}, and 
explaining the origin and survival of particles at these energies would require either new physics or a nearby source 
of $\gtrsim 10^{20}$ eV cosmic rays \citep{Bertolami2000}. Upper limits for the cosmic ray flux at these energies is 
severely detector limited, and the probability of detecting events $>10^{20}$ eV becomes vanishingly small. 

Are there cosmic rays of energy $\gg 10^{20}$ eV, and if so, how can they be detected? Possible indirect detection 
strategies for cosmic rays over a wide range of energies involve gamma ray observations of e.g. the galactic 
center \citep{Su2010}, regions of dark matter \citep{Sciama2000}, and molecular clouds \citep{Casanova2010}.

Atmospheres of planets and stars could themselves act as large detectors for indirect detection of very energetic cosmic 
ray events. The cosmic ray induced gamma ray emission in Earth's atmosphere has been mapped by \citet{Petry2005} 
and has recently been measured by Fermi in order to constrain the cosmic ray spectrum between 90 GeV and 
6 TeV \citep{Fermi2014}. Searching for energetic cosmic ray events in the atmosphere of our Sun is difficult because the 
Sun itself is the source of energetic particles as well as gamma radiation \citep[e.g.][]{Kanbach1993}. Nevertheless, 
quiescent solar gamma radiation is detected by the Fermi satellite, and is attributed to cosmic 
rays (\citealt{Orlando2009}, using the model of \citealt{Seckel1991}).

Jupiter is a natural candidate for remote detection of extensive air showers, because it is a large dense object that is 
not expected to produce its own gamma ray emission. Jupiter has been seen in in auroral emission, due to a complex 
interplay between volcanic particulates from Io, the Jovian magnetosphere and the solar 
wind \citep[see, eg.][]{Kim1998}. There is a tentative gamma ray detection from Jupiter, setting a flux limit 
of $6.3 \times 10^{-7}$ erg cm$^{-2}$ s$^{-1}$ \citep{Fichtel1975}. The cause of this emission is unknown. 
\citet{dePater2004} has measured the radio synchrotron variability on Jupiter with LOFAR, and this variability 
is similar to the variability observed in brown dwarfs \citep{Heinze2013}.

In this paper, we explore the feasibility of treating Jupiter as a detector of galactic cosmic ray events of energy 
$E_0 > 10^{20}$ eV. In Section \ref{sec:frequency} we estimate how often these cosmic ray events can occur 
within Jupiter's atmosphere. We discuss the electromagnetic component of the shower in a model atmosphere of Jupiter 
in Section \ref{sec:atmosphere-Jupiter}. In Section \ref{sec:observation}, we discuss gamma ray emission 
(Sect. \ref{sec:gamma}) and synchrotron emission (Sect. \ref{sec:synchrotron}) from events occurring within 
Jupiter's atmosphere. Section \ref{sec:conclusion} contains our concluding remarks on the feasibility of detecting these 
signatures.

\section{Occurrence of Extensive Air Showers}
\label{sec:frequency}
In order to explore the frequency of extensive air showers on Jupiter, we first need to determine the integrated flux of 
cosmic rays of primary energy $E_0 > 10^{18}$ eV, called Ultra-High-Energy Cosmic Rays (UHECRs). The flux-density for 
primary cosmic rays of $E_0 < 4 \times 10^{19}$ eV can be fitted by 
$j(E_0) \propto E_0^{-2.7}$ \citep[][their Eq. 26.2]{Beringer2012}. There is evidence that, above this energy, the 
assumed power-law becomes steeper \citep{Abraham2010b}. We therefore use a broken power-law spectrum kinked at 
$E_{10} = 4 \times 10^{10}$ GeV:
\begin{equation}
j_p(E_0) \approx  j_0 
\begin{cases}
E_0^{-\epsilon}, & E_0 < E_{10} \\
E_{10}^{\kappa - \epsilon} \, E_0^{-\kappa}, & E_0 > E_{10}
\end{cases}
\label{eqn:crflux}
\end{equation}
where $j_0 = 1.8$ cm$^{-2}$ s$^{-1}$ sr$^{-1}$ GeV$^{-1}$, $\epsilon = 2.7$, $\kappa = 4.2$, and $[E_0] = {\rm GeV}$.

We can use this flux density to determine a frequency of extensive air showers above a given energy. Since we are 
interested in the highest energy cosmic rays, we calculate the frequency for extensive air shower events over a 
``detector'' with surface area $A\Omega$ [cm$^2$ sr], $\nu_{\rm EAS}$ [s], expressed as:
\begin{equation}
 \nu_{\rm EAS} = A\Omega \int_{E_{0}}^{\infty} j_p(E) \; dE.
 \label{eqn:event-timescale}
\end{equation}
For the Pierre Auger Observatory, with a surface area of roughly $A\Omega \approx 2\pi \, 3000$ km$^2$, the average 
amount of time between two $4 \times 10^{19}$ eV events is about 1.5 days. For Jupiter, $A\Omega \approx 4\pi R_J^2$, 
where $R_J = 6.99 \times 10^9$ cm is Jupiter's radius, and the time between $4 \times 10^{19}$ eV events is about 0.1 
seconds. We will not be able to observe the direct consequences of all extensive air showers occurring in Jupiter's 
atmosphere. The effective detector area is calculated for gamma ray observations in Section \ref{sec:gamma}. This 
effective area will depend on integrated column densities through Jupiter's atmosphere, and therefore on the 
temperature profile of Jupiter.

\section{Atmospheric and Extensive Air Shower Model for Jupiter}
\label{sec:atmosphere-Jupiter}

We now apply Heitler's approximation for an extensive air shower \citep{Matthews2005} to a model of Jupiter's atmosphere.
We use the one-dimensional temperature profile for the 30$^{\circ}$N Latitude vernal equinox model for Jupiter 
from \citet{Moses2005}, and we make the assumption that Jupiter's temperature-profile is isotropic. The profile 
from \citet{Moses2005} was constructed mostly from observations, at pressures $> 10^{-3}$ bar from infrared observations 
from ISO and at pressures $< 10^{-6}$ bar from the Galileo probe. Between $10^{-6}$ bar and $10^{-3}$ 
bar, \citet{Moses2005} assume isothermal temperatures. This temperature profile and bulk chemistry have
an effect on the development of extensive air showers.

The penetration depth for an extensive air shower is parameterized by the shower age, $s = s(X)$, such that $s = 0$ 
corresponds to the initiating event, $s = 1$ corresponds to the shower maximum, and $s \rightarrow 3$ as 
$X \rightarrow \infty$. The relationship between $s$ and $X$ depends on various atmospheric properties. The shower age 
can be fit with a broken power law \citep{Abraham2010c}:
\begin{equation}
 s(X) = 
\begin{cases} 
 \dfrac{3}{1 + \dfrac{2 \chi_0}{X}\log\Big(\dfrac{E_0}{E_{\rm crit}}\Big)}, & E_0 \leq E_9;\\
  \dfrac{3}{1 + \dfrac{2 \chi_0'}{X}\Big|\log\Big(\dfrac{E_0}{E_{\rm crit}}\Big)\Big|^{0.35}}, & E_0 > E_9;
 \end{cases} 
 \label{eqn:shower-age}
\end{equation}
where $E_9 = 2\times 10^{18}$ eV, $\chi_0$ is the radiation path-length and $\chi_0'$ is given its value such that $s(X)$ 
is continuous over $E_9$. The values of the parameters $r_M$, $E_{\rm crit}$ and $\chi_0$ are taken from the Particle 
Data Group, \url{pdg.lbl.gov}, with the atmospheric chemical and physical properties taken from \citet{Moses2005}. 
For this model atmosphere, $E_{\rm crit} = 3.448 \times 10^8$ eV is the critical energy, 
$\chi_0 = 63.04$ g cm$^{-2}$ and $\chi_0' = 277.2$ g cm$^{-2}$ are characteristic path-lengths.

The number of electrons in an extensive air shower, $N_e$, has been related to the energy of the initiating cosmic ray 
by \citet[][their Eq. 2.3]{Nagano1992}. We use Heitler's approximation for the number of electrons at the shower maximum 
\citep{Matthews2005}:
\begin{equation}
 N_e \approx \dfrac{E_0}{10 \, E_{\rm crit}}
 \label{eqn:electron-number}
\end{equation}
The normalized energy distribution of secondary electrons is independent of the energy and composition of the initial 
cosmic ray, according to CORSIKA simulations \citep{Nerling2006}. The analytic function for the distribution 
is \citep[][their Eq. 9]{Nerling2006}:
\begin{equation}
 \eta(s,E) = \dfrac{k_0E \; e^{k_1s - k_2 s^2}}{\big(E + 6.43 - 1.53s\big)\big(E + 168. - 42.1s\big)^s},
 \label{eqn:eta-sE}
\end{equation}
where the parameters $k_0 = 0.142$, $k_1 = 6.180$ and $k_2 = 0.606$, and $s(X)$ is the shower age. We use these 
calculations for hypothetical air showers in Jupiter's atmosphere in order to estimate their gamma ray and synchrotron 
signatures.

\section{Direct Observation of Extensive Air Showers on Jupiter}
\label{sec:observation}

We explore the feasibility of observing extensive air showers in Jupiter's atmosphere by gamma ray emission 
(Section \ref{sec:gamma}) and Synchrotron emission (Section \ref{sec:synchrotron}).

\subsection{Gamma Ray Emission}
\label{sec:gamma}

\begin{figure}
\centering
\includegraphics[width=\columnwidth]{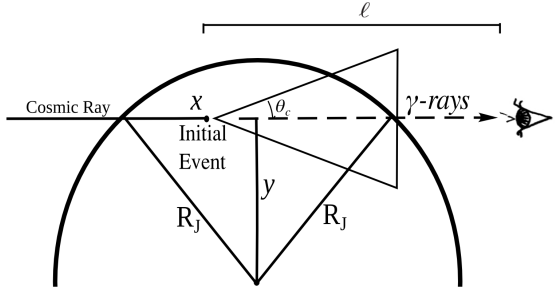}
\caption{Diagram of a cosmic ray initiating an extensive air shower and producing photons on a trajectory toward an 
observer on Earth. Cartesian coordinates are oriented such that the $x$-axis is along the path of the cosmic
ray and extensive air shower. $R_J = 6.9911 \times 10^9$ cm denotes the radius of Jupiter.}
\label{fig:geometry}
\end{figure}

Extensive air showers begin with $\pi^+$, $\pi^-$ and $\pi^0$ particles. The $\pi^0$ particle decays quickly into 
$2\gamma$. After traveling a characteristic column on the order of the radiation length, the $\gamma$-rays will 
catastrophically produce $e^+e^-$ pairs. These electrons in turn produce photons via Bremsstrahlung 
emission \citep{Gaisser1990}. We adopt the approximation of \citet{Matthews2005} and estimate that the number of
photons from an electromagnetic sub-shower is $\sim 10\times$ the number of electrons, keeping in mind that the flux spectrum produced 
by Bremsstrahlung is fairly flat up to the energy of the electron. We therefore make the crude approximation that photons spanning
an energy of $\Delta E_{\gamma}$ are of number $N_{\gamma} \approx N_e(E_e)$ for all $E_{\gamma} < E_e$. We can determine
$N_e(E_e)$ from Eq.'s (\ref{eqn:electron-number}) and (\ref{eqn:eta-sE}). These photons are spread over an area the size of 
$\pi (\ell \tan \theta_c)^2$, where $\ell$ [cm] is the distance through the atmosphere between the initiating event for the air shower 
and the observer, which we take to be $\sim 5$ AU, and $\theta_c$ is the opening angle of the shower (Fig. \ref{fig:geometry}). 
The statistical root mean square of the angle between the axis of the shower and the photons is 
$\theta_c \approx m_ec^2/E_e$ \citep{Gaisser1990}. The gamma ray fluence, $\Phi$, for an extensive air shower 
is approximately:
\begin{equation}
\Phi \approx \int\limits_{\Delta E_{\gamma}} \dfrac{N_{\gamma}(E_{\gamma}) \; dE_{\gamma}}{\pi \ell^2 \tan^2 \big(m_ec^2/E_e\big)}, 
\label{eq:fluence}
\end{equation}
where the integral is taken over the range of energies relevant for a given gamma ray detector, $\Delta E_{\gamma}$ [eV].
Because the fluence depends on the square of $\tan \theta_c$, which for small angles is $\sim (E_e/m_ec^2)^2$,
 the highest energy electrons in the distribution (Eq. (\ref{eqn:eta-sE})) will determine the maximum fluence. The maximum
fluence is only observed if the center of the gamma ray shower strikes the detector.
The gamma rays produced by the highest energy electrons are effectively pencil beams, and so it is more likely to observe the
gamma ray component at less than the maximum fluence. We will now consider the effective ``detector area'' for gamma ray emission from 
extensive air showers in Jupiter's atmosphere, and then will return to the question of whether Fermi-LAT can detect these showers.

Some extensive air showers will be directed toward the Earth, and will not have traveled though so much of Jupiter's 
atmosphere as to be significantly attenuated. If Jupiter is thought of as a giant cosmic ray detector, we can use 
this fact to estimate the effective detector area that Jupiter provides for observing cosmic rays. In order to make
this estimate, we need to consider cosmic rays transiting through a ring in the upper atmosphere of Jupiter bookended by 
two column densities. For column densities below $\sim 30$ g cm$^{-2}$, it is unlikely that UHECRs will experience a 
collision, and so the cosmic rays will stream through the atmosphere unaffected. For column densities much 
above $\sim 710$ g cm$^{-2}$, the gamma ray signature from an extensive air shower will be attenuated by the 
atmosphere \citep{Alvarez2002,Beringer2012}. Extensive air showers will therefore be remotely observable at a range 
of column densities $30$ g cm$^{-2}$ $\lesssim X_{\perp} \lesssim 710$ g cm$^{-2}$. Above $710$ g cm$^{-2}$, we assume complete
attenuation, and below $710$ g cm$^{-2}$ we assume no attenuation (i.e. we assume the best-case scenario,
that the shower maximum is achieved just before the shower exits the atmosphere of Jupiter). We calculate $X_{\perp}$ using the 
density profile for Jupiter from \citet{Moses2005} as follows.

\begin{figure}
\centering
\includegraphics[width=\columnwidth]{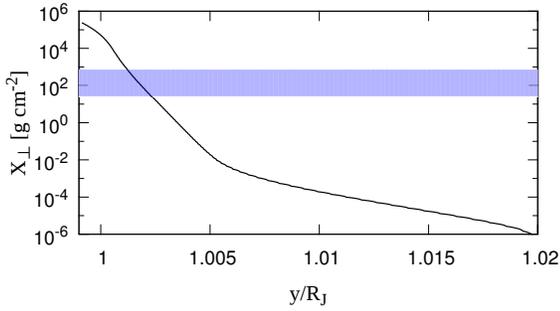}
\caption{The integrated column density, $X_{\perp}$ [g cm$^{-2}$] versus the ratio $y/R_J$. The blue shaded region 
represents the range of column densities where gamma rays from cosmic ray air showers could in principle be detected.}
\label{fig:projection}
\end{figure}

We can solve for the column density experienced by a cosmic ray  penetrating into Jupiter, with its closest approach to 
the center of Jupiter a distance $y$ [cm]. A diagram of the cosmic ray path is shown in Figure \ref{fig:geometry}. The 
column density is obtained by solving the integral:
\begin{equation}
 X_{\perp}(y) = \!\!\!\! \int\limits_{-\sqrt{R_J^2-y^2}}^{\sqrt{R_J^2-y^2}} \!\!\!\! dx \; n\big(h(x,y)\big).
\end{equation}
where $n(h)$ [cm$^{-3}$] is the number density in Jupiter's atmosphere at a hight $h(x,y) = \sqrt{x^2 + y^2}$.
The solution of this integral for various values of $y$ is plotted in Figure \ref{fig:projection}. The effective 
``detector'' for the Jovian atmosphere is a ring with a surface area of $A\Omega \approx 2rR_J\Omega$, where 
$r \approx 10^{-3}R_J$. The solid angle, $\Omega$, through which the cosmic ray must travel in order to produce an air 
shower that Fermi can detect is restricted by the opening angle, $\theta_c$. The angular spread of extensive air 
showers occurring in the region $30$ g cm$^{-2}$ $\lesssim X_{\perp} \lesssim 710$ g cm$^{-2}$ projects a ring at a 
distance $\ell$ from Jupiter with surface area $A\Omega \approx 4\pi \ell R_J m_e c^2/E_e$.

We now compare the gamma ray fluence from an event in Jupiter's atmosphere, $\Phi$, to the detection limit for 
gamma ray fluence for Fermi's Gamma-ray Burst Monitor (GBM) and Fermi LAT using the sensitivity estimates and the 
spectral range from \citet{Meegan2009} and \citet{Atwood2009}, respectively. Fermi-GBM observes gamma rays in an
energy range of $\Delta E_{\gamma} = 0.008-10$ MeV and a sensitivity of $\sim 10^{-5}$ erg cm$^{-2}$, and Fermi-LAT with 
$\Delta E_{\gamma} = 0.02-300$ GeV and a sensitivity of $\sim 10^{-7}$ erg cm$^{-2}$. At its highest fluence sensitivity,
Fermi-LAT can achieve an angular resolution of $\sim 0.5^{\circ}$.

\begin{figure}
\centering
\includegraphics[trim=9mm 0mm 15mm 0mm, clip, width=\columnwidth]{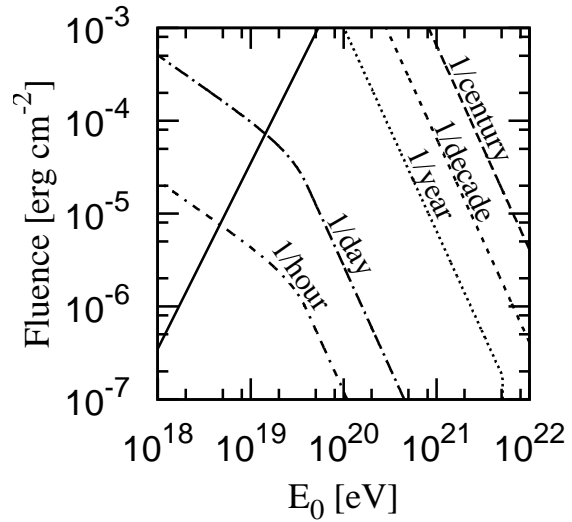}
\caption{A contour plot of cosmic ray frequency as a function of the energy of the extensive air-shower, $E_0$ [eV], and 
the gamma ray fluence [erg cm$^{-2}$] as observed by Fermi-LAT. The solid line indicates the maximum fluence 
from an extensive air shower of energy $E_0$.}
\label{fig:crfrequency}
\end{figure}

The detection limits for gamma ray fluence 
can be applied to Eq. (\ref{eq:fluence}) in order to solve for $E_e$. The effective cosmic ray detector area for Jupiter, 
as observed by Fermi, is:
\begin{equation}
 A\Omega = \dfrac{2}{5}\Big(\dfrac{R_J\Delta E_{\gamma}}{\ell \Phi}\Big)\Big(\dfrac{E_0}{m_ec^2}\Big) \leq 4\pi r R_J.
\label{eqn:surface-area}
\end{equation}
We now solve Eq. (\ref{eqn:event-timescale}), using Eq. (\ref{eqn:surface-area}) for $A\Omega$. Our resulting cosmic ray 
detection frequency for Fermi-LAT is presented as a function of the gamma ray fluence, $\Phi$, and the energy of 
the extensive air shower, in Fig. \ref{fig:crfrequency}. We predict that Fermi-LAT should be able to detect events of 
energy $>10^{21}$ eV with fluence $10^{-7}$ erg cm$^{-2}$ at a rate of about one per month. Due to its lower energy 
range, we do not predict that Fermi-GBM will be able to detect any extensive air showers from Jupiter.

\subsection{Synchrotron Emission}
\label{sec:synchrotron}

The electronic component of an extensive air shower would have a characteristic spectroscopic signature due to 
synchrotron emission. We can calculate an approximate power-spectrum for the synchrotron emission from the relativistic 
electrons within our extensive air shower.

\begin{figure}
\centering
\includegraphics[width=\columnwidth]{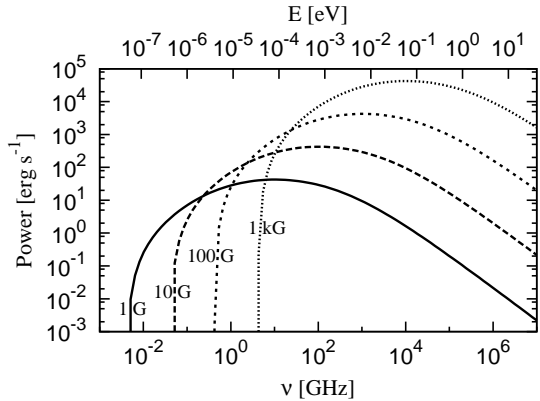}
\caption{The synchrotron emission power [erg s$^{-1}$] for extensive air showers of energy $10^{18}$ eV, 
at four different magnetic field strengths, ranging from 1 Gauss to 1 kGauss, as a function of 
frequency $\nu$ [s$^{-1}$].}
\label{fig:Pfig}
\end{figure}

\begin{figure}
\centering
\includegraphics[width=\columnwidth]{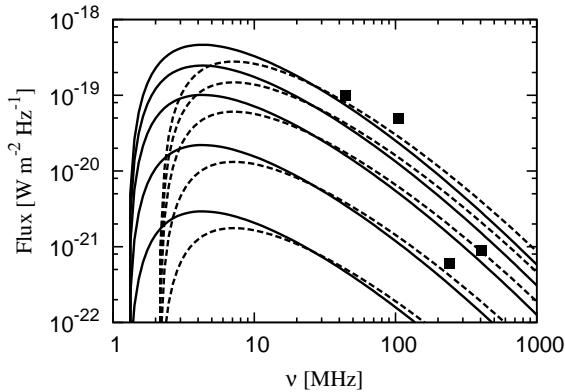}
\caption{The synchrotron emission flux [W m$^{-2}$ Hz$^{-1}$] for extensive air showers of energy $10^{16} - 10^{17}$ eV,
as a function of frequency, $\nu$ [s$^{-1}$] for magnetic field strengths of 0.3 Gauss (solid lines) and 0.5 Gauss 
(dashed lines). These spectra are compared to the data from \citet[][black points]{Spencer1969}.}
\label{fig:EarthP}
\end{figure}

Imagine a number of electrons, $\mathcal{N}_{\gamma}$, each in the same magnetic field, $B$, and each with the same 
velocity characterized by $\gamma = (1 - v^2/c^2)^{-1/2}$. These electrons will all radiate at an identical synchrotron 
frequency \citep[][their Eq. 29]{Felten1966}:
\begin{equation}
 \nu \; {\rm [s^{-1}]} \approx \dfrac{3e\gamma^2B}{4\pi m_e c},
 \label{eqn:sync-frequency}
\end{equation}
and will radiate with a power \citep[][their Eq. 32]{Felten1966}:
\begin{equation}
 P \; {\rm [erg \, s^{-1}]} = \dfrac{2\gamma^2e^4B^2\mathcal{N}_{\gamma}^2}{3m_e^2c^3}.
 \label{eqn:sync-power}
\end{equation}
Given a distribution of electrons, $\eta(E)$, from Eq. (\ref{eqn:eta-sE}), we can determine the power as a function of 
the frequency. The relativistic factor can be given as a function of the energy, $\gamma = E/(m_ec^2) + 1$, and we can 
give the electron energy in terms of the synchrotron frequency produced by electrons of that energy:
\begin{equation}
 E(\nu) = m_ec^2 \Bigg( \sqrt{\dfrac{4\pi m_e c \nu}{3eB}} - 1 \Bigg).
 \label{eqn:E-nu}
\end{equation}
The number of free electrons that will produce a synchrotron signature at the frequency $\nu$ is then given by: 
$\mathcal{N}_{\gamma} = N_e \eta(\nu)$. For an extensive air shower, $N_e$ from Eq. (\ref{eqn:electron-number}) and 
$\eta(\nu)$ comes from Eq. (\ref{eqn:eta-sE}) with the energy from Eq. (\ref{eqn:E-nu}). The radiation power is then:
\begin{equation}
 P(\nu) = \dfrac{8\pi e^3 B N_e^2}{9 m_ec^2} \, \nu \eta^2\!\big(E(\nu)\big).
\label{eqn:power}
\end{equation}
The resulting synchrotron spectrum for extensive air showers is plotted as a function of $\nu$ in Figure \ref{fig:Pfig}.
Our results compare well with both the synthetic spectrum and radio observations presented in \citet{Falcke2003}, if the 
ambient magnetic field is set to 0.3 Gauss. \citet{Falcke2003} fit their synthetic spectrum (their Fig. 2) to the data
from \citet{Spencer1969}. The data can be well-fit by cosmic rays of energies $10^{16}$ eV - $10^{17}$ eV. Our comparison 
with the data from \citet{Spencer1969} for magnetic field strengths of 0.3-0.5 Gauss is plotted in 
Figure \ref{fig:EarthP}. This comparison demonstrates the variation of the synchrotron power spectrum with cosmic ray 
energy.

Jupiter's magnetic field of $B_J \sim 4$ G will shift the peak frequency of the synchrotron emission to approximately 
6 mm, bringing the synchrotron signature into the detection range of the Atacama Large Millimeter/submillimeter Array. 
Although it may be virtually impossible to disentangle the synchrotron emission from extensive air showers from 
other microwave emission on Jupiter, coincident detections could be achieved by simultaneously observing Jupiter with a 
millimeter telescope and the Fermi LAT.

Since the peak frequency of the synchrotron emission depends on the energy distribution of the electrons, and this 
distribution is expected to be independent of the cosmic ray energy \citep{Nerling2006}, the peak frequency of this 
emission will depend on the local magnetic field. If the local magnetic field where the extensive air showers occurs 
is different from $B_J$, then the peak frequency will shift as shown in Figure \ref{fig:Pfig}. Radio and infrared 
synchrotron emission from events on Jupiter is therefore also possible.

\section{Concluding Remarks}
\label{sec:conclusion}

We suggest observing Jupiter for signatures of very high energy cosmic ray events, first by considering how often
extensive air showers occur on the surface of Jupiter, and then by examining the gamma ray and synchrotron signatures 
of such an air shower. If gamma ray observations could be connected to extensive air showers in Jupiter's atmosphere, 
this would afford the cosmic ray community a significantly larger detector area and could help constrain the cosmic ray 
spectrum above $10^{20}$ eV. We can apply Jupiter's ``detector'' cross-section of $3 \times 10^{7}$ km$^2$ to 
Eq. (\ref{eqn:event-timescale}) to estimate an expected cosmic ray event frequency. If the observed number of events
significantly exceeded one every $12$ hours, this would be evidence of GZK violation, where cosmic rays survive at
energies greater than those at which interaction with the blueshifted cosmic microwave background becomes significant.
This would imply either a local source of $>10^{20}$ eV cosmic rays or new physics. We suggest using the
Fermi observatory to observe this gamma ray signature.

Brown dwarfs are comparable to Jupiter in size. Since brown dwarfs are far more distant than Jupiter, gamma 
ray detection of extensive air showers in their atmospheres would be unlikely in the foreseeable future. 
Nevertheless, it may be possible to observe the synchrotron emission from the electrons comprising extensive air showers, 
and the occurrence of extensive air-showers in atmospheres of free-floating substellar objects, far from any host star, 
may be an important source of electrification within their cloud layers. We will examine these possibilities
in a later paper.

 \subsection*{Acknowledgement}
All authors highlight financial support of the European Community under the FP7 by an ERC starting grant. P.B.R. is 
grateful for critical corrections from Alan Watson on an earlier draft, Alexander MacKinnon for helpful discussions, and 
Ian Taylor for his technological support. We thank the anonymous referee for their helpful comments.
This research has made use of NASA's Astrophysics Data System.

\bibliographystyle{apj}

\end{document}